\documentclass{acm_proc_article-sp}

\pdfoutput=1

\usepackage{cite}
\usepackage{graphicx}
\usepackage{url}
\usepackage{listings}

\lstdefinestyle{FortranLike}{float,frame=lines,language=Fortran,commentstyle=\ttfamily,basicstyle=\ttfamily}
\lstdefinestyle{CLike}{float,frame=lines,language=C,commentstyle=\ttfamily,basicstyle=\ttfamily}
\lstdefinestyle{NoFloatCLike}{frame=lines,language=C,commentstyle=\ttfamily,basicstyle=\ttfamily}

\begin{document}

\title{ForOpenCL: Transformations Exploiting Array Syntax in Fortran for Accelerator Programming}

\numberofauthors{6}

\author{
\alignauthor
Matthew J. Sottile\\
       \affaddr{Galois, Inc.}\\
       \affaddr{Portland, OR 97204}\\
       \email{mjsottile@computer.org}
\alignauthor
Craig E Rasmussen\\
       \affaddr{Los Alamos National Laboratory}\\
       \affaddr{Los Alamos, NM 87545}\\
       \email{crasmussen@lanl.gov}
\alignauthor
Wayne N. Weseloh\\
       \affaddr{Los Alamos National Laboratory}\\
       \email{weseloh@lanl.gov}
\and  
\alignauthor
Robert W. Robey\\
       \affaddr{Los Alamos National Laboratory}\\
       \email{brobey@lanl.gov}
\alignauthor
Daniel Quinlan\\
       \affaddr{Lawrence Livermore National Laboratory}\\
       \email{dquinlan@llnl.gov}
\alignauthor
Jeffrey Overbey\\
       \affaddr{University of Illinois at Urbanna-Champaign}\\
       \email{overbey2@illinois.edu}
}

\maketitle

\begin{abstract}
  Emerging GPU architectures for high performance computing are well suited to a
  data-parallel programming model.  This paper presents preliminary work
  examining a programming methodology that provides Fortran programmers with access
  to these emerging systems.  We use array constructs in Fortran to
  show how this infrequently exploited, standardized language feature is easily
  transformed to lower-level accelerator code.  The transformations in ForOpenCL are based
  on a simple mapping from Fortran to OpenCL.  We demonstrate, using a
  stencil code solving the shallow-water fluid equations, that the performance
  of the ForOpenCL compiler-generated transformations is comparable with that of
  hand-optimized OpenCL code.
\end{abstract}


\section{Introduction}
\label{sec:intro}

This paper presents a compiler-level approach for targeting a single program to
multiple, and possibly fundamentally different, processor architectures.  This
technique allows the application programmer to adopt a single, high-level
programming model without sacrificing performance.  We suggest that existing
data-parallel features in Fortran are well suited to applying automatic
transformations that generate code specifically tuned for different hardware
architectures using low-level programming models such as OpenCL.  For algorithms
that can be easily expressed in terms of whole array, data-parallel operations,
writing code in Fortran and transforming it automatically to specific low-level
implementations removes the burden of creating and maintaining multiple versions
of architecture specific code.

The peak performance of these newer accelerator architectures can be substantial.
Intel expects a teraflop for the SGEMM benchmark with their Knights Ferry
processor while the performance of the M2090 NVIDIA Tesla processor is in the
same neighborhood~\cite{hpcwire11manycore}.  Unfortunately the performance that
many of the new accelerator architectures offer comes at a cost.  Architectural
changes are trending toward multiple heterogeneous cores and less of a reliance
on superscalar instruction level parallelism and hardware managed memory
hierarchies (such as traditional caches).  These changes place a heavy burden on
application programmers as they work to adapt to these new systems.  An
especially challenging problem is not only how to program to these new
architectures --- considering the massive scale of concurrency available --- but
also how to design programs that are portable across the changing landscape of
computer architectures.  How does a programmer write one program that can
perform well on both a conventional multicore CPU \emph{and} a GPU (or another
of the emerging many-core architectures)?

A directive-based approach, such as OpenMP or the Accelerator programming model
from the Portland Group~\cite{pgi10accelerator}, is one solution to this
problem.  However, in this paper we take a somewhat different approach.  A
common theme amongst the new processors is the emphasis on data-parallel
programming.  This model is well suited to architectures that are
based on either vector processing or massively parallel collections of simple
cores.  The recent CUDA and OpenCL programming languages are intended to
support this programming model.

The problem with OpenCL and CUDA is that they expose too much detail about the
machine architecture to the programmer~\cite{wolfe08gpgpu}.  The programmer is
responsible for explicitly managing memory (including the staging of data back
and forth between the host CPU and the accelerator device) and specifically
taking into account architectural differences (such as whether the architecture
contains vector units).  While these languages have been attractive as a method
for early adopters to utilize these new architectures, they are less attractive
to programmers who do not have the time or resources to manually port their code
to every new architecture and programming model that emerges.

\subsection{Approach}

We demonstrate that a subset of Fortran map surprisingly well onto GPUs when
transformed to OpenCL kernels.  This data-parallel subset includes: array syntax
using assignment statements and binary operators, array constructs like {\tt
  WHERE}, and the use of pure and elemental functions.  In addition, we provide
new functions that explicitly take advantage of the stencil geometry of the
problem domain we consider.  Note that this subset of the Fortran language is
implicitly parallel.  This programming model \emph{does not require explicit
  declaration of parallelism within the program.}  In addition, programs are
expressed using entirely standard Fortran so it can be compiled for and executed
on a single core without concurrency.

Transformations are supplied that provide a mechanism for converting Fortran
procedures written in the Fortran subset described in this paper to OpenCL
kernels.  We use the ROSE compiler
infrastructure\footnote{\url{http://www.rosecompiler.org/}} to develop these
transformations.  ROSE uses the Open Fortran
Parser\footnote{\url{http://fortran-parser.sf.net/}} to parse Fortran 2008
syntax and can generate C-based OpenCL.  Since ROSE's intermediate
representation (IR) was constructed to represent multiple languages, it is
relatively straightforward to transform high-level Fortran IR nodes to C OpenCL
nodes.  This work is also applicable to transformations to vendor-specific
languages, similar to OpenCL, such as the NVIDIA CUDA language.

Transformations for arbitrary Fortran procedures are not attempted.
Furthermore, a mechanism to transform the calling site to automatically invoke
OpenCL kernels is not provided at this time.  While it is possible to accomplish
this task within ROSE, it is considered outside the scope of this paper.
However, ForOpenCL provides via Fortran interfaces a mechanism to call the C OpenCL
runtime and enable Fortran programmers to access OpenCL kernels generated by the
supplied transformations.

We study the automatic transformations for an application example that is
typical of stencil codes that update array elements according to a fixed
pattern.  Stencil codes are often employed in applications based on
finite-difference or finite-volume methods in computational fluid dynamics
(CFD).  The example described later in this paper is a simple shallow-water
model in two dimensions using finite volume methods.

Finally, we examine the performance of the Fortran data-parallel abstraction
when transformed to OpenCL to run on GPU architectures.  The performance of
automatically transformed code is compared with a hand-optimized OpenCL
version of the shallow-water code.

\section{Programming Model}

A question that one may pose is ``\emph{Why choose Fortran and not a
  more modern language like X for programming accelerator
  architectures?}''  The recent rise in interest in concurrency and
parallelism at the language level driven by multicore CPUs and
many-core accelerators has driven a number of new language
developments, both as novel languages and extensions on existing ones.
However, for many scientific users with existing codes written in
Fortran, new languages and language extensions to use novel new
architectures present a challenge: how do programmers effectively use
them while avoiding rewriting code and potentially growing dependent
on a transient technology that will vanish tomorrow?  In this paper we
explore the constructs in Fortran that are particularly relevant to
GPU architectures.

In this section we present the Fortran subset employed in this paper.
This sub-setting language will allow scientific programmers to stay
within the Fortran language and yet have direct access to GPU
hardware.  We start by examining how this programming model relates to
developments in other languages.

\subsection{Comparison to Prior Fortran Work}

A number of previous efforts have exploited data-parallel programming
at the language level to utilize novel architectures.  The origin of
the array syntax adopted by Fortran in the 1990 standard can be found
in the APL language~\cite{iverson79apl}.  These additions to Fortran
allowed parallelism to be expressed with whole-array operations at the
expression level, instead of via parallelism within explicit DO-loops,
as implemented in earlier variants of the language (e.g., IVTRAN for the
Illiac IV).

The High Performance Fortran (HPF) extension of Fortran was proposed
to add features to the language that would enhance the ability of
compilers to emit fast parallel code for distributed and shared memory
parallel computers\cite{koelbel94hpf}.  One of the notable additions
to the language in HPF was syntax to specify the distribution of data
structures amongst a set of parallel processors.  HPF also introduced
an alternative looping construct to the traditional DO-loop called
{\tt FORALL} that was better suited for parallel compilation.  An
additional keyword, {\tt INDEPENDENT}, was added to allow the
programmer to indicate when the loop contained no loop-order
dependencies that allowed for parallel execution.  These constructs
are similar to {\tt DO CONCURRENT}, an addition to Fortran in 2008.

Interestingly, the parallelism features introduced in HPF did not
exploit the new array features introduced in 1990 in any significant
way, relying instead on explicit loop-based parallelism.  This
restriction allowed the language to support parallel programming that
wasn't easily mapped onto a pure data-parallel model.  The {\tt SHADOW}
directive introduced in HPF-2, and the {\tt HALO} in HPF+~\cite{benkner99hpf}
bear some similarity to the halo region concept that we discuss in this 
paper.

In some instances though, a purely data-parallel model is appropriate
for part or all of the major computations within a program.  One of
the systems where programmers relied heavily on higher level
operations instead of explicit looping constructs was the Thinking
Machines Connection Machine 5 (CM-5).  A common programming pattern
used on the CM-5 (that we exploit in this paper) was to write
whole-array operations from a global perspective in which computations
are expressed in terms of operations over the entire array instead of
a single local index.  The use of the array shift intrinsic functions
(like {\tt CSHIFT}) were used to build computations in which arrays
were combined by shifting the entire arrays instead of working on
local offsets based on single indices.  A simple 1D example is one in
which an element is replaced with the average of its own value and
that of its two direct neighbors.  Ignoring boundary indices that wrap
around, explicit indexing will result in a loop such as:

{\small
\begin{verbatim}
  do i = 2,(n-1)
    X(i) = (X(i-1) + X(i) + X(i+1)) / 3
  end do
\end{verbatim}
}

\noindent When shifts are employed, this can be expressed as:

{\small
\begin{verbatim}
  X = (cshift(X,-1) + X + cshift(X,1)) / 3
\end{verbatim}
}

\noindent
Similar whole array shifting was used in higher dimensions for finite difference
codes within the computational physics community for codes targeting the CM-5
system.  Research in compilation of stencil-based codes that use shift operators
targeting these systems is related to the work presented
here~\cite{stencil-compiler}.

The whole-array model was attractive because it deferred responsibility for
optimally implementing the computations to the compiler.  Instead of relying on
a compiler to infer parallelism from a set of explicit loops, the choice for how
to implement loops was left entirely up to the tool.

Unfortunately, this had two side effects that have limited broad
acceptance of the whole-array programming model in Fortran.  First,
programmers must translate their algorithms into a set of global
operations.  Finite difference stencils and similar computations are
traditionally defined in terms of offsets from some central index.
Shifting, while conceptually analogous, can be awkward to think about
for high dimensional stencils with many points.  Second, the semantics
of these operations are such that all elements of an array operation
are updated as if they were updated simultaneously.  In a program
where the programmer explicitly manages arrays and loops, double
buffering techniques and user managed temporaries are used to maintain
these semantics.  Limited attention to optimizing memory usage due to
this intermediate storage by compilers has led to these constructs seeing
little adoption by programmers.


An interesting line of language research that grew out of HPF was that
associated with the ZPL language at the University of
Washington~\cite{chamberlain04zpl} and Chapel, an HPCS language developed by
Cray~\cite{chamberlain_chapel}.  In ZPL, programmers adopt a similar global view
of computation over arrays, but define their computations based on regions,
which provide a local view of the set of indices that participate in the update
of each element of an array.  A similar line of research in the functional
language community has investigated array abstractions for expressing
whole-array operations in the Haskell language in the REPA (regular, shape-polymorphic,
parallel array) library~\cite{keller10repa}.

\subsection{Fortran Language Subset}

The static analysis and source-to-source transformations used in this work
require the programmer to use a language subset that employs a data-parallel
programming model.  In particular, it encourages the use of array notation,
elemental functions, and pure procedures.  From these language constructs, we
are able to easily transform Fortran procedures to a lower-level OpenCL kernel
implementation.

\subsubsection*{Array notation}

Fortran has a rich array syntax that allows programmers to write statements in
terms of whole arrays or subarrays, with data-parallel operators to compute on
the arrays.  Array variables can be used in expressions based on whole-array
operations.  For example, if {\tt A}, {\tt B}, and {\tt C} are all arrays of the
same rank and shape and {\tt s} is a scalar, then the statement

{\small
\begin{verbatim}
 C = A + s*B
\end{verbatim}
}

\noindent
results in the element-wise sum of {\tt A} and the product of {\tt s} times the
elements of {\tt B} being stored in the corresponding elements of {\tt C}. The
first element of {\tt C} will contain the value of the first element of {\tt A}
added to the first element of {\tt c*B}.  Note that no explicit iteration over
array indices is needed and that the individual operators, plus, times, and
assignment are applied by the compiler to individual elements of the arrays
independently.  Thus the compiler is able to spread the computation in the
example across any hardware threads under its control.

\subsubsection*{Elemental functions}

An elemental function consumes and produces scalar values, but can be applied to
variables of array type such that the function is applied to each and every
element of the array.  This allows programmers to avoid explicit looping and
instead simply state that they intend a function to be applied to every element
of an array in parallel, deferring the choice of implementation technique to the
compiler.  Elemental functions are intended to be used for data-parallel
programming, and as a result must be side effect free and mandate an {\tt
intent(in)} attribute for all arguments.

For example, the basic array operation shown above could be refactored into an
elemental function,

{\small
\begin{verbatim}
  pure elemental real function foo(a, b, s)
    real, intent(in) :: a, b, s
    foo = a + s*b
  end function
\end{verbatim}
}

\noindent and called with

{\small
\begin{verbatim}
  C = foo(A, B, s)
\end{verbatim}
}

Note that while {\tt foo} is defined in terms of purely scalar quantities, it
can be \emph{applied} to arrays as shown.  While this may seem like a trivial
example, such simple functions may be composed with other elemental functions to
perform powerful computations, especially when applied to arrays.  Our prototype
tool transforms elemental functions to inline OpenCL functions.  Thus there is
no penalty for usage of elemental functions and they provide a convenient
mechanism to express algorithms in simpler segments.

\subsubsection*{Pure procedures}

Pure procedures, like elemental functions, must be free of side effects.  Unlike
elemental functions that require arguments to have an {\tt intent(in)}
attribute, they may change the contents of array arguments that are passed to
them.  The absence of side effects removes ordering constraints that could
restrict the freedom of the compiler to invoke pure functions out of order and
possibly in parallel.  Procedures and functions of this sort are also common in
pure functional languages like Haskell, and are exploited by compilers in order
to emit parallel code automatically due to their suitability for compiler-level
analysis.

Since pure procedures don't have side effects they are candidates for running on
accelerators in OpenCL.  Currently our prototype tool only transforms pure procedures
to OpenCL kernels that \emph{do not} call other procedures, except for elemental
functions, either defined by the user or intrinsic to Fortran.

\subsection{New Procedures}

Borrowing ideas from ZPL, we introduce a concept of a region to Fortran with a
set of functions that allow programmers to work with subarrays in expressions.
In Fortran, these functions return a copy of or a pointer to an existing array or array section.
This is unlike ZPL, where regions are analogous to index sets and are used
primarily for address resolution within an array without dictating storage
related behavior.  The functions that we propose are similar in that they allow
a programmer to deal with index regions that are meaningful to their algorithm,
and automatically induce a halo (or ghost) cell pattern as needed in the
implementation generated by the compiler, where the size of an array is
implicitly increased to provide extra array elements surrounding the interior 
portion of the array.  It is important to note, however, that all memory allocated
by the programmer must explicitly contain the extra array elements in the halo.

Region functions are similar to the shift operator as they can be used to
reference portions of the array that are shifted with respect to the interior
portion.  However, unlike the shift operator, regions are not expressed in terms
of boundary conditions and thus don't explicitly \emph{require} a knowledge of,
nor the application of, boundary conditions locally (global boundary conditions must
be explicitly provided by the programmer outside of calls to kernel procedures).
Thus, as will be shown
below, regions are more suitable for usage by OpenCL thread groups which access
only local subsections of an array stored in global memory.

ForOpenCL provides two new functions that are defined in Fortran and are used
in array-syntax operations.  Each function takes an integer array halo argument
that specifies the number of ghost cells on either side of a region, for each
dimension.  For example {\tt halo = [left, right, down, up]} specifies a halo
for a two-dimensional region.  These functions are:

\begin{itemize}

\item {\tt region\_cpy(array, halo)}: a pure function that returns a copy of
  the interior portion of the array specified by halo.

\item {\tt region\_ptr(array, halo)}: an impure function that returns a pointer 
  to the portion of the array specified by halo.

\end{itemize}

It should be noted that the function {\tt region\_cpy} is pure and thus can be
called from within a pure kernel procedure, though it should be noted that {\tt
region\_ptr} is impure because it aliases the array parameter.  However as will
be shown below, the usage of {\tt region\_ptr} is constrained so that it does not
introduce side effects in the functions that call it.  These two functions are
part of the language recognized by the compiler and though {\tt region\_cpy}
returns a copy of a portion of an array \emph{semantically}, the compiler is not
forced to actually make a copy and is free to enforce copy semantics through
other means.  In addition to these two new functions, ForOpenCL provides the
compiler directive, {\tt \!\$OFP PURE, KERNEL}, which specifies that a pure subroutine
can be transformed to an OpenCL kernel and that the subroutine is pure
except for calls to {\tt region\_ptr}.  These directives are not strictly
necessary for the technique described in this paper, but aid in automated
identification of specific kernels to be transformed to OpenCL.  A directive-free
implementation would require the transformation tool be provided the set of 
kernels to work via a user defined list.

\subsection{Advantages}

There are several advantages to this style of programming using array
syntax, regions, and pure and elemental functions:

\begin{itemize}
\item There are no loops or index variables to keep track of.  Off by
  one index errors and improper handling of array boundaries are a
  common programming mistake.
\item The written code is closer to the algorithm, easier to
  understand, and is usually substantially shorter.
\item Semantically the intrinsic function {\tt region\_cpy} returns an array by value.
  This is usually what the algorithm requires.
\item Pure and elemental functions are free from side effects, so it is
  easier for a compiler to schedule the work to be done in parallel.
\end{itemize}

Data parallelism has been called collection-oriented programming by
Blelloch~\cite{blelloch90}.  As the {\tt cshift} function and the array-valued
expressions all semantically return a value, this style of programming is also
similar to functional programming (or value-oriented programming).  It should be
noted that the sub-setting language we employ goes beyond pure data parallelism
by the use of pure (other than calls to {\tt region\_ptr}) subroutines and
not just elemental functions.

Unfortunately, this style of programming has never really caught on because when
array syntax was first introduced in Fortran, performance of codes using these features was
relatively poor and thus programmers shied away from using array syntax (even
recently, some are actively counseling against its usage because of performance
issues~\cite{Levesque:SC08}).  Thus the Fortran community was caught in a
classic ``chicken-and-egg'' conundrum: (1) programmers didn't use it because it
was slow; and (2) compilers vendors didn't improve it because programmers didn't
use it.  A goal of this paper is to demonstrate that parallel programs written
in this style of Fortran can achieve good performance on accelerator architectures.

\subsection{Restrictions}

Only pure Fortran procedures are transformed into OpenCL kernels.  This restriction
is lifted slightly to allow calls to {\tt region\_ptr} from within a kernel procedure.  The
programmer must explicitly call these kernels using Fortran interfaces in the ForOpenCL
library (described below).  It is also possible, using ROSE, to modify the calling
site so that the entire program can be transformed, but this functionality is
outside the scope of this paper.  Here we specifically examine transforming
Fortran procedures to OpenCL kernels.  Because OpenCL support is relatively new
to ROSE, some generated code must be modified.  For example, the {\tt
  \_\_global} attribute for kernel arguments was added by hand.

It is assumed that memory for all arrays reside on the device.  The programmer
must copy memory to and from the device.  In addition, array size (neglecting
ghost cell regions) must be multiples of the global OpenCL kernel size.

Array variables within a kernel procedure (specified by the {\tt \!\$OFP PURE,
  KERNEL} directive) must be declared as contiguous.  A kernel procedure may not
call other procedures except for limited intrinsic functions (primarily math),
user-defined elemental functions, and the {\tt region\_cpy} and {\tt region\_ptr}
functions.  Future work will address non-contiguous arrays (such as those that
result from strided access) by mapping array strides to gather/scatter-style
memory accessors.

Array parameters to a kernel procedure must be declared as either intent(in) or
intent(out); they cannot be intent(inout).  A thread may read from an extended
region about its local element (using the {\tt region\_cpy} function), but can
only write to the single array element it owns.  If a variable were intent(inout),
a thread could update its array element before another thread had read from that
element.  This restriction requires double buffering techniques.

\section{Shallow Water Model}
\label{sec:shallow-water}

The numerical code used for this work is from a presentation at the
NM Supercomputing Challenge~\cite{Robey07}.
The algorithm solves the standard 2D shallow water equations. This
algorithm is typical of a wide range of modeling equations based on
conservation laws such as compressible fluid dynamics (CFD), elastic
material waves, acoustics, electromagnetic waves and even traffic
flow~\cite{Leveque02}. For the shallow water problem there are
three equations with one based on conservation of mass and the other
two on conservation of momentum.

\begin{eqnarray*}
h_{t}+(hu)_{x}+(hv)_{y} & = & 0\quad\mbox{(mass)}\\
(hu)_{t}+(h{u}^{2}+\tfrac{1}{2}gh^{2})_{x}+(huv)_{y} & = & 0\mbox{\quad($x$-momentum)}\\
(hv)_{t}+(huv)_{x}+(h{v}^{2}+\tfrac{1}{2}gh^{2})_{y} & = & 0\mbox{\quad($y$-momentum)}
\end{eqnarray*}


\noindent
where h = height of water column (mass), $u$ = x velocity, $v$ =
y velocity, and $g$ = gravity. The height $h$ can be used for mass
because of the simplification of a unit cell size and a uniform water
density. Another simplifying assumption is that the water depth is
small in comparison to length and width and so velocities in the z-direction
can be ignored. A fixed time step is used for simplicity though it
must be less than $dt \leqq dx / (\sqrt{gh}+|u|)$ to fulfill the CFL
condition.

The numerical method is a two-step Lax-Wendroff scheme. The method
has some numerical oscillations with sharp gradients but is adequate
for simulating smooth shallow-water flows. In the following explanation,
$U$ is the conserved state variable at the center of the cell. This
state variable, $U$ $=(h,hu,hv)$ in the first term in the equations
below. $F$ is the flux quantity that crosses the boundary of the cell
and is subtracted from one cell and added to the other. The remaining
terms after the first term are the flux terms in the equations above
with one term for the flux in the x-direction and the next term for
the flux in the y-direction. The first step estimates the values a
half-step advanced in time and space on each face, using loops on
the faces.\begin{eqnarray*}
U_{i+\frac{1}{2},j}^{n+\frac{1}{2}} & = & (U_{i+1,j}^{n}+U_{i,j}^{n})/2+\frac{\triangle t}{2\triangle x}\left(F_{i+1,j}^{n}-F_{i,j}^{n}\right)\\
U_{i,j+\frac{1}{2}}^{n+\frac{1}{2}} & = & (U_{i,j+1}^{n}+U_{i,j}^{n})/2+\frac{\triangle t}{2\triangle y}\left(F_{i,j+1}^{n}-F_{i,j}^{n}\right)\end{eqnarray*}

The second step uses the estimated values from step 1 to compute the
values at the next time step in a dimensionally unsplit loop.\[
U_{i,j}^{n+1}=U_{i,j}^{n}-\frac{\triangle t}{\triangle x}(F_{i+\frac{1}{2},j}^{n+\frac{1}{2}}-F_{i-\frac{1}{2},j}^{n+\frac{1}{2}})-\frac{\triangle t}{\triangle y}(F_{i,j+\frac{1}{2}}^{n+\frac{1}{2}}-F_{i,j-\frac{1}{2}}^{n+\frac{1}{2}})\]

\subsection{Fortran implementation}

Selected portions of the data-parallel implementation of the shallow water model
are now shown.  This code serves as input to the ForOpenCL transformations
described in the next section.  The interface for the Fortran kernel procedure {\tt
  wave\_advance} is declared as:

{\small
\begin{verbatim}
subroutine wave_advance(dx,dy,dt,H,U,V,oH,oU,oV)
  !$OFP PURE, KERNEL   :: wave_advance
  real, intent(in)     :: dx,dy,dt
  real, dimension(:,:) :: H,U,V,oH,oU,oV
  contiguous  :: H,U,V,oH,oU,oV
  intent(in)  :: H,U,V
  intent(out) :: oH,oU,oV
  target      :: oH,oU,oV
end subroutine
\end{verbatim}
}

\noindent
where {\tt dx, dy, dt} are differential quantities in space $x, y$ and time $t$,
{\tt H, U}, and {\tt V} are state variables for the height and $x$ and $y$
momentum respectively, and {\tt oH, oU, oV} are corresponding output arrays used
in the double buffering scheme.  The \emph{OFP} compiler-directive attributes
{\tt PURE} and {\tt KERNEL} indicate that the procedure {\tt wave\_advance} is
to be transformed as an OpenCL kernel and that it must be pure, other than for any
pointers used to reference interior regions of the output arrays.

Temporary arrays are required for the quantities {\tt Hx, Hy, Ux, Vx}, and {\tt
  Vy,} that are defined on cell faces.  Also, the pointer variables, {\tt pH,
  pU,} and {\tt pV}, are needed to access and update interior regions of the
output arrays.  As these pointers are assigned to the arrays {\tt oH, oU, oV},
these output arrays must have the target attribute, as shown in the interface
above.  The temporary arrays and array pointers are declared as,

{\small
\begin{verbatim}
real, allocatable, dimension(:,:) :: Hx, Hy, Ux
real, allocatable, dimension(:,:) :: Uy, Vx, Vy
real, pointer,     dimension(:,:) :: pH, pU, pV
\end{verbatim}
}

Halo variables for the interior and the cell faces are declared and defined as

{\small
\begin{verbatim}
integer, dimension(4) :: face_lt, face_rt, halo
integer, dimension(4) :: face_up, face_dn

halo    = [1,1,1,1]
face_lt = [0,1,1,1];  face_rt = [1,0,1,1]
face_dn = [1,1,0,1];  face_up = [1,1,1,0]
\end{verbatim}
}

\noindent
Note that the halo definitions for the four faces each have a 0 in the
initialization. Thus the returned array copy will have a size that is larger
than any interior region that uses the full halo {\tt [1,1,1,1]}.  This is
because there is one more cell face quantity than there are cells in a given
direction.

The first Lax-Wendroff step updates state variables on the cell faces.  Assignment
statements like the following,

{\small
\begin{verbatim}
Hx = 0.5*( region_cpy(H,face_lt) +   &
           region_cpy(H,face_rt) )   &
     + (0.5*dt/dx)                   &
     * (region_cpy(U,face_lt) - region_cpy(U,face_rt))
\end{verbatim}
}

\noindent
are used to calculate these quantities.  This equation updates the array for the
height in the $x$-direction.  The second step then uses these face quantities to
update the interior region, for example,

{\small
\begin{verbatim}
face_lt = [0,1,0,0];  face_rt = [1,0,0,0]
face_dn = [0,0,0,1];  face_up = [0,0,1,0]

pH = region_ptr(oH, halo)

pH = region_cpy(H, halo)
       + (dt/dx) * ( region_cpy(Ux, face_lt) -   &
                     region_cpy(Ux, face_rt) )   &
       + (dt/dy) * ( region_cpy(Vy, face_dn) -   &
                     region_cpy(Vy, face_up) )
\end{verbatim}
}

\noindent
Note that face halos have been redefined so that the array copy
returned has the same size as the interior region.

These simple code segments show how the shallow water model is implemented in
standard Fortran using the data-parallel programming model described above.  The
resulting code is simple, concise, and easy to understand.  However it does
\emph{not} necessarily perform well when compiled for a traditional sequential
system because of suboptimal use of temporary array variables,
especially those produced by the function {\tt region\_cpy}.  This is generally true
of algorithms that use Fortran shift functions as well, as some Fortran
compilers (e.g., gfortran) do not generate optimal code for shifts.  We note (as
shown below) that these temporary array copies are replaced by scalars in the
transformed Fortran code so there are no performance penalties for using
data-parallel statements as outlined.  However, there is an increased memory
cost due to the double buffering required by the kernel execution semantics.

\section{Source-To-Source Transformations}

This section provides an brief overview of the ForOpenCL
transformations that take Fortran elemental and pure procedures as
input and generate OpenCL code.  Elemental functions are
transformed to inline OpenCL functions and subroutines with the
{\tt PURE} and {\tt KERNEL} compiler directive attributes are
transformed to OpenCL kernels.

\subsection{OpenCL}

OpenCL~\cite{opencl08} is an open-language standard for developing applications
targeted for GPUs, as well as for multi-threaded applications targeted for
multi-core CPUs.  The kernels are run by calling a C runtime library from the
OpenCL host (normally the CPU).  Efforts to standardize a C++ runtime are
underway and Fortran interfaces to the C runtime are distributed in the ForOpenCL
library.

An important concept in OpenCL is that of a thread and a thread group.  Thread
groups are used to run an OpenCL kernel concurrently on several processor
elements on the OpenCL device (often a GPU).  Consider a data-parallel statement
written in terms of an elemental function as discussed above.  The act of
running an OpenCL kernel can be thought of as having a particular thread
assigned to each instance of the call to the elemental function as it is mapped
across the arrays in the data-parallel statement.  In practice, these threads
are packaged into thread groups when they are run on the device hardware.

Device memory is separated hierarchically.  A thread instance has access to its
own thread memory (normally a set of registers), threads in a thread group to
OpenCL local memory, and all thread groups have access to OpenCL global memory.
When multiple members of a thread group access the same memory elements (for
example in the use of the {\tt region\_cpy} function in the calculation of face
variable quantities shown above), for performance reasons it is often best that
global memory accessed and shared by a thread group be copied into local memory.

\begin{figure}[!t]
\centering
\includegraphics[width=1.75in]{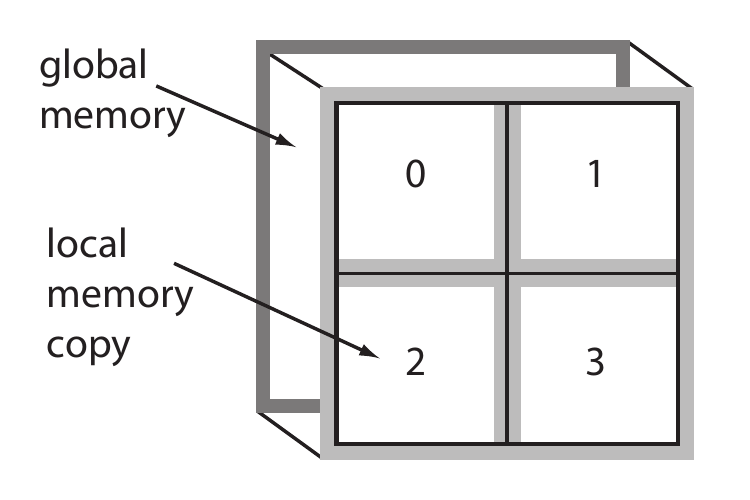}
\caption{A schematic of global memory for an array and its copy stored in local memory
for four thread groups.}
\label{fig:cl-memory}
\end{figure}

The \emph{region} and \emph{halo} constructs easily map onto the OpenCL memory
hierarchy.  A schematic of this mapping is shown in Figure~\ref{fig:cl-memory}
for a two-dimensional array with a 2x2 array of 4 thread groups.  The memory
for the array and its halo are stored in global memory on the device as shown
in the background layer of the figure.  The array copy in local memory is
shown in the foreground divided into 4 \emph{local} tiles that partition the
array.  Halo regions in global memory are shown in dark gray and halo regions
in local memory are shown in light gray.

We point out that the hierarchical distribution of memory used on the OpenCL
device shown in Figure~\ref{fig:cl-memory} is similar to the distribution of
memory across MPI nodes in an MPI application.  In the case of MPI, the
virtual global array is represented by the background layer (with its halo) and
partitions of the global array are stored in the 4 MPI nodes shown in the foreground.
Our current and future work on this effort includes source-to-source transformations
to generate MPI code in addition to OpenCL in order to deal with clusters of 
nodes containing accelerators.  This work is outside the scope of this paper.

Halo regions obtained via the {\tt region\_cpy} function (used with intent(in)
arrays) are constrained semantically so that they can not be written to by an
OpenCL kernel.  The {\tt region\_cpy} function returns a copy of the region of
the array stored in global memory and places it in local memory shared by
threads in a thread group.  Thus once memory for an array has been transferred
into global device memory by the host (before the OpenCL kernel is run), memory
is in a consistent state so that all kernel threads are free to read from global
device memory.  Because the local memory is a copy, it functions as a software
cache for the local thread group.  Thus the compiler must insert OpenCL barriers
at proper locations in the code to insure that all threads have written to the
local memory cache before any thread can start to read from the cache.  On exit
from a kernel, any local memory explicitly stored in register variables by the
compiler (memory accessed via the {\tt region\_cpy} function) is copied back to
global memory for all intent(out) arrays.  Recall that a thread may only write to
its own intent(out) array element, thus there are no race conditions when updating
intent(out) arrays.




\subsection{Transformation examples}

This section outlines the OpenCL equivalent syntax for portions of the Fortran
shallow-water code described in Section~\ref{sec:shallow-water}.  The notation
uses uppercase for arrays and lowercase for scalar quantities.  Variables
temporarily storing quantities for updated output arrays (declared as pointers
in Fortran) are denoted by a {\tt p} preceding the array name.  For example, the
Fortran statement {\tt pH = region\_ptr(oH, halo)} is transformed as a scalar
variable declaration representing a single element in the output array {\tt oH}.

\vskip .4in

\subsubsection{Region function}

While the Fortran version of the {\tt region\_cpy} function semantically returns
an array copy, in OpenCL this function returns a scalar quantity based on the
location of a thread in a thread group and the relationship of its location to
the array copy transferred to local memory.  Because we assume there is a thread
for every element in the interior, the array index is just the thread index
adjusted for the size of the halo.  Thus {\tt region\_cpy} is just an inline
OpenCL function and is provided by the ForOpenCL library.

\subsubsection{Function and variable declarations}

Fortran kernel procedures have direct correspondence with OpenCL equivalents.
For example, the {\tt wave\_advance} interface declaration is transformed as
{\small
\begin{verbatim}
__kernel void
wave_advance(float dx, ..., __global float * H, ...);
\end{verbatim}
}
\noindent
The intent(in) arrays have local equivalents that are stored in local memory
and are declared by, for example,
{\small
\begin{verbatim}
__local float H_local[LOCAL_SIZE];
\end{verbatim}
}
\noindent
These local arrays are declared with the appropriate size and are
copied to local memory by the compiler with an inlined library function.  The
array temporaries defined on cell faces are declared similarly while interior
pointer variables are simple scalars, e.g., {\tt float pH}.  Intent(in) array
variables cannot be scalar objects because regions may be shifted and thus
\emph{shared} by threads within a thread group.

\subsubsection{Array syntax}

Array syntax transforms nearly directly to OpenCL code.  For example, interior
pointer variables are particularly straightforward as they are scalar quantities in
OpenCL,
{\small
\begin{verbatim}
pH = region_cpy(H, halo)
       + (dt/dx) * ( region_cpy(Ux, face_lt) -
                     region_cpy(Ux, face_rt) )
       + (dt/dy) * ( region_cpy(Vy, face_dn) -
                     region_cpy(Vy, face_up) );
\end{verbatim}
}
\noindent
Allocated variables are more complicated because they are arrays.
{\small
\begin{verbatim}
Hx[i] = 0.5 * (region(H_local, face_lt)+ ...);
\end{verbatim}
}
\noindent
where {\tt i = LX + LY*(NLX+halo(0)+halo(1)))} is a local index variable, {\tt
  LX = get\_local\_id(0)} is the local thread id in the $x$ dimension, {\tt LY =
  get\_local\_id(1)} is the local thread id in the $y$ dimension, {\tt NLX =
  get\_local\_size(0)} is the size of the thread group in the $x$ dimension, and
the {\tt get\_local\_id} and {\tt get\_local\_size} functions are defined by the
OpenCL language standard.

\section{Performance Measurements}

Performance measurements were made comparing the transformed code with different
versions of the serial shallow-water code.  The serial versions included two
separate Fortran versions: one using data-parallel notation and the other using
explicit looping constructs.  We also compared with a hand-written OpenCL
implementation that was optimized for local memory usage (no array temporaries).
The accelerated measurements were made using an NVIDIA Tesla C2050 (Fermi) cGPU
with 2.625 GB GDDR5 memory, and 448 processor cores.  The serial measurements
were made using an Intel Xeon X5650 hexacore CPU with 96 GB of RAM running at
2.67 GHz.  The compilers were gfortran and gcc version 4.4.3 with an
optimization level of -O3.




Several timing measurements were made by varying the size of the array state
variables.  The performance measurements are shown in Table~\ref{table:performance}.  An average time was obtained by executing 100 iterations of the outer
time-advance loop that called the OpenCL kernel. This tight loop kept the OpenCL
kernel supplied with threads to take advantage of potential latency hiding by
the NVIDIA GPU.  Any serial code within this loop (not present in this study)
would have reduced the measured values.

\begin{table}
\begin{center}
	\begin{tabular}{|c|c|c|c|}
	\hline Array width & F90 & GPU (16x8) & Speedup \\ \hline\hline
	16   & 0.025 ms & 0.017 ms & 1.5  \\ \hline
	32   & 0.086    & 0.02     & 4.3  \\ \hline
	64   & 0.20     & 0.02     & 10.0 \\ \hline
	128  & 0.76     & 0.036    & 21.1 \\ \hline
	256  & 3.02     & 0.092    & 32.8 \\ \hline
	512  & 12.1     & 0.32     & 37.8 \\ \hline
	1024 & 49.5     & 1.22     & 40.6 \\ \hline
	1280 & 77.7     & 1.89     & 41.1 \\ \hline
	2048 & 199.1    &  4.82    & 41.3 \\ \hline
	4096 & 794.7    & 19.29    & 41.2 \\ \hline
	\end{tabular}
\end{center}
\caption{Performance measurements for the shallow-water code.  All times reported in milliseconds.}
\label{table:performance}
\end{table}

The transformed code achieved very good results.  In all instances, the
performance of the transformed code was within 25\% of the hand-optimized OpenCL
kernel.  Most of the extra performance of the hand-optimized code can be
attributed to the absence of array temporaries and to packing the three state
variables {\tt H, U}, and {\tt V} into a single vector datatype.

While we did not have an OpenMP code for multi-core comparisons, the transformed
OpenCL code on the NVIDIA C2050 was up to 40 times faster than the best serial
Fortran code executing on the host CPU.


\section{Conclusions}

The sheer complexity of programming for clusters of many or multi-core
processors with tens of millions threads of execution makes the simplicity of
the data-parallel model attractive.  The increasing complexity of
todays applications (especially in light of the increasing complexity
of the hardware) and the need for portability across architectures
make a higher-level and simpler programming model like data-parallel
attractive.

The goal of this work has been to exploit source-to-source transformations that
allow programmers to develop and maintain programs at a high-level of
abstraction, without coding to a specific hardware architecture.
Furthermore these transformations allow multiple hardware architectures
to be targeted without changing the high-level source.  It also removes the
necessity for application programmers to understand details of the accelerator
architecture or to know OpenCL.

\section{Acknowledgments}
This work was supported in part by the Department of Energy Office of Science,
Advanced Scientific Computing Research.


\bibliographystyle{abbrv}
\bibliography{foropencl}

\balancecolumns
\end{document}